\documentclass[10pt,twocolumn,letterpaper]{article}

\usepackage{cvpr}
\usepackage{times}
\usepackage{epsfig}
\usepackage{graphicx}
\usepackage{amsmath}
\usepackage{amssymb}
\usepackage{xcolor}
\usepackage{algorithm} 
\usepackage{algpseudocode} 
\usepackage{paralist}
\usepackage{enumitem}
\usepackage{subcaption}

\DeclareMathOperator*{\argmin}{arg\,min}

\usepackage[pagebackref=true,breaklinks=true,letterpaper=true,colorlinks,bookmarks=false]{hyperref}

\cvprfinalcopy 

\def\cvprPaperID{****} 

\ifcvprfinal\fi
\begin{document}

\newcommand{\hly}{\colorbox{yellow}}
\newcommand{\hlg}{\colorbox{green}}
\newcommand{\hlp}{\colorbox{pink}}

\title{Underwater Image Super-Resolution using Deep Residual Multipliers}

\author{ Md Jahidul Islam, Sadman Sakib Enan, Peigen Luo, and Junaed Sattar \\	
{\tt\small \{islam034, enan0001, luo00034, junaed\}@umn.edu} \\
{\small Interactive Robotics and Vision Laboratory, Department of Computer Science and Engineering} \\ 
{\small Minnesota Robotics Institute, University of Minnesota, Twin Cities, MN, USA }
}

\maketitle

\begin{abstract}
We present a deep residual network-based generative model for single image super-resolution (SISR) of underwater imagery for use by autonomous underwater robots. We also provide an adversarial training pipeline for learning SISR from paired data. In order to supervise the training, we formulate an objective function that evaluates the \textit{perceptual quality} of an image based on its global content, color, and local style information. Additionally, we present USR-248, a large-scale dataset of three sets of underwater images of `high' $(640\times480)$ and `low' $(80\times60$, $160\times120$, and $320\times240)$ spatial resolution. USR-248 contains paired instances for supervised training of $2\times$, $4\times$, or $8\times$ SISR models. Furthermore, we validate the effectiveness of our proposed model through qualitative and quantitative experiments and compare the results with several state-of-the-art models' performances. We also analyze its practical feasibility for applications such as scene understanding and attention modeling in noisy visual conditions.
\end{abstract}

\section{Introduction}
Visually-guided autonomous underwater vehicles require image synthesis and scene understanding in many important applications such as the monitoring of marine species and coral reefs~\cite{hoegh2007coral}, inspection of submarine cables and wreckage~\cite{bingham2010robotic}, human-robot collaboration~\cite{islam2018understanding}, and more.
Autonomous Underwater Vehicles (AUVs) and Remotely Operated Vehicles (ROVs) are widely used in these applications, where they harness the synthesized images for visual attention modeling to make navigation decisions such as `where to look or go next', `which snapshots should be recorded', etc. 
However, despite often using high-end cameras, underwater images are often greatly affected~\cite{islam2019fast} by poor visibility, absorption, and scattering. 
Consequently, the objects of interest may appear blurred as the images lack important details. 
This problem exacerbates when the camera (\ie, robot) cannot get close to the objects to get a closer view, \eg, while following a fast-moving target, or surveying distant coral reefs or seabed. Fast and accurate techniques for Single Image Super-Resolution (SISR) can alleviate these problems by restoring the perceptual and statistical qualities of the low-resolution image patches. 

\begin{figure}[t]
	\centering
	\begin{subfigure}{0.5\textwidth}
		\includegraphics[width=\linewidth]{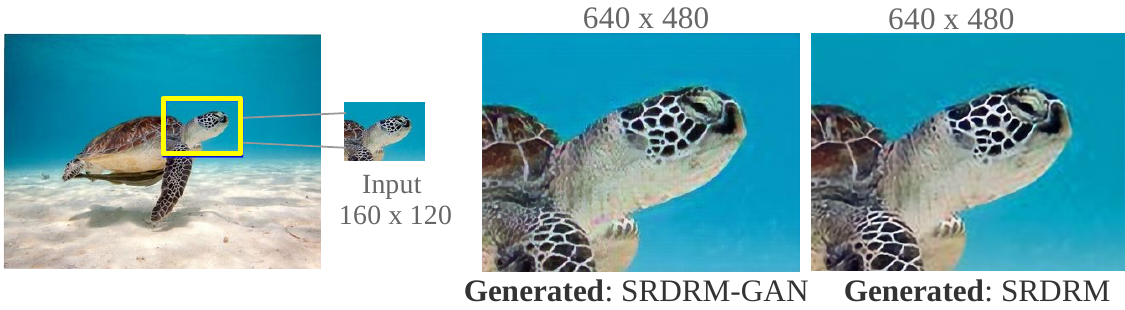} 
		\caption{Zoom-in capability: HR image generation from LR image patches}
	\end{subfigure}
	
	\vspace{1mm}
	\begin{subfigure}{0.5\textwidth}
		\includegraphics[width=\linewidth]{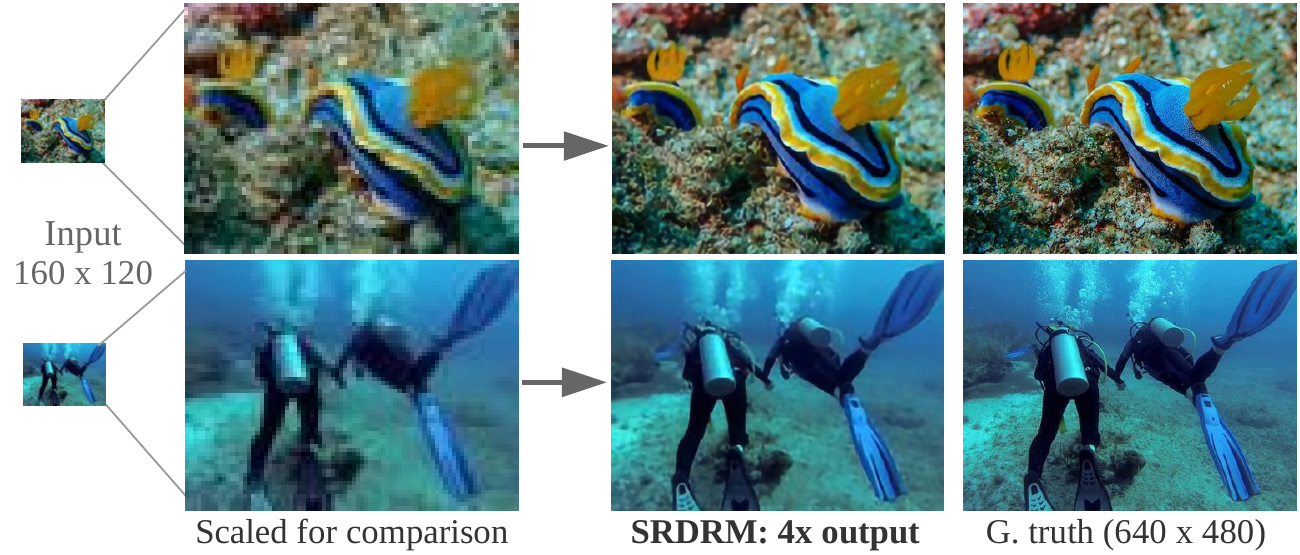} 
		\caption{Realistic HR image generation: comparison with the ground truths}
	\end{subfigure}
	
	\caption{Demonstration of underwater image super-resolution using our proposed models: SRDRM and SRDRM-GAN.}
	\label{fig:1}
\end{figure}

The existing literature based on deep Convolutional Neural Networks (CNNs) provides good solutions for automatic SISR~\cite{dong2015image,lim2017enhanced}.  
In particular, several Generative Adversarial Network (GAN)-based models provide state-of-the-art (SOTA) performance~\cite{wang2018esrgan,ledig2017photo} in learning to enhance image resolution from a large collection of paired or unpaired data~\cite{yuan2018unsupervised}. 
However, there are a few challenges involved in adopting such models for underwater imagery. First, the underwater images suffer from a set of unique distortions. For instance, they tend to have a dominating green or blue hue because the red wavelengths get absorbed in deep water~\cite{fabbri2018enhancing}. Other factors such as the lighting variations in different depths, amount of particles in the water, and scattering cause irregular non-linear distortions which result in low-contrast and blurry images~\cite{islam2019fast}.
Consequently, the off-the-shelf SISR models trained on arbitrary images fail to generate realistic higher resolution underwater images.
Secondly, the lack of large-scale underwater dataset restricts extensive research attempts for the training and performance evaluation of SISR models on underwater images. 
Because of the high costs and difficulties associated with acquiring real-world underwater data, the existing datasets (that were originally proposed for training object detection and image enhancement models) often contain synthetic images~\cite{fabbri2018enhancing} and/or their resolution are typically limited to $256\times256$~\cite{islam2019fast}. 
Due to these challenges, designing SISR models for underwater imagery and investigating their applicability in real-world underwater robotic applications have not been explored in-depth in the literature.

We attempt to address these challenges by designing a novel SISR model that can learn to generate $2\times$, $4\times$, or $8\times$ higher resolution (HR) underwater images from the respective low-resolution (LR) inputs. 
We also present a large-scale underwater dataset that provides the three sets of LR-HR pairs of images 
used to train the proposed model.     
In addition, we perform thorough experimental evaluations of the proposed model and demonstrate its effectiveness compared to several existing SOTA models. 
Specifically, we make the following contributions in this paper: 
\begin{compactenum}[(a)]

\vspace{1mm}
\item We present a fully-convolutional deep residual network-based generative model for underwater SISR, which we refer to as \textbf{SRDRM}. 
We also formulate an adversarial training pipeline (\ie, \textbf{SRDM-GAN}) by designing a multi-modal objective function that evaluates the perceptual image quality based on its global content, color, and local style information. In our implementation, SRDRM and SRDM-GAN can learn to generate $640\times480$ images from respective inputs of size $320\times240$, $80\times60$, or $160\times120$. The model and associated training pipelines are available at \url{https://github.com/xahidbuffon/srdrm}.   

\vspace{1mm}
\item In addition, we present \textbf{USR-248}, a collection of over $1050$ samples (\ie, paired HR-LR images) that facilitate large-scale SISR training. It has another $248$ test images for benchmark evaluation.  
These images are rigorously collected during oceanic explorations and field experiments, and also from a few publicly available online resources. We make this available at \url{http://irvlab.cs.umn.edu/resources/usr-248-dataset}.

\vspace{1mm}
\item Furthermore, we perform a number of qualitative and quantitative experiments that validate that the proposed model can learn to enhance underwater image resolution from both traditional and adversarial training.  
We also analyze its feasibility and effectiveness for improving visual perception in underwater robotic applications; a few sample demonstrations are highlighted in Fig.~\ref{fig:1}.  

\end{compactenum}

\section{Related Work}\label{related_work}

\subsection{Single Image Super-resolution (SISR)}
SISR has been studied~\cite{freeman2002example, chang2004super, melville2005super} for nearly two decades in the area of signal processing and computer vision. 
Some of the classical SISR methods include statistical methods~\cite{sun2008image,kim2010single,protter2008generalizing}, patch-based methods~\cite{glasner2009super,yang2012coupled, huang2015single}, sparse representation-based methods~\cite{yang2010image}, random forest-based method \cite{schulter2015fast}, etc. 
In recent years, with the rapid development of deep learning-based techniques, this area of research has been making incredible progress. In the pioneering work, Dong \textit{et al.}~\cite{dong2015image} proposed a three-layer CNN-based end-to-end model named SRCNN, that can learn a non-linear LR-HR mapping without requiring any hand-crafted features. Soon after, Johnson \textit{et al.}~\cite{johnson2016perceptual} showed that replacing the per-pixel loss with a perceptual loss (that quantifies image quality) gives better results for CNN-based SISR models. 
On the other hand, Kim \textit{et al.} proposed deeper networks such as VDSR~\cite{kim2016accurate},~DRCN \cite{kim2016deeply} and used contemporary techniques such as gradient clipping, skip connection, and recursive-supervision in order to improve the training further. 
Moreover, the sparse coding-based networks~\cite{liu2016robust}, residual block-based networks (\eg, EDSR~\cite{lim2017enhanced}, DRRN~\cite{tai2017image}), and other CNN-based models~\cite{shi2016real},~\cite{dong2016accelerating} have been proposed that outperform SRCNN for SISR. 
These methods, however, have rather complex training pipelines, and are often prone to poor performance for large scaling factors (\ie, 4$\times$ and higher). 
Thus far, researchers have been trying to address these issues by using Laplacian pyramid-based networks (LapSRN)~\cite{lai2017deep}, 
dense skip connections (SRDenseNet)~\cite{tong2017image}, deep residual networks (RDN)~\cite{zhang2018residual}, etc. 

The CNN-based SISR models learn a sequence of non-linear filters from a large number of training images. This end-to-end learning of LR-HR mapping provide significantly better performance~\cite{yang2019deep} compared to using hand-crafted filters, or traditional methods based on bicubic interpolation. 
On the other hand, Generative Adversarial Networks (GANs)~\cite{goodfellow2014generative} employ a two-player min-max game where the `generator' tries to fool the `discriminator' by generating \textit{fake} images that appear to be \textit{real} (\ie, sampled from the HR distribution). Simultaneously, the discriminator tries to get better at discarding fake images and eventually (in equilibrium) the generator learns the underlying LR-HR mapping. 
GANs are known to provide SOTA performance for style transfer~\cite{gatys2016image} and image-to-image translation~\cite{isola2017image} problems in general.  
As for SISR, the GAN-based models can recover finer texture details~\cite{sonderby2016amortised,chen2018efficient} while super-resolving at large up-scaling factors. 
For instance, Ledig \textit{et al.} showed that SRGAN~\cite{ledig2017photo} can reconstruct high-frequency details for an up-scaling factor of $4$. 
Moreover, ESRGAN~\cite{wang2018esrgan} incorporates a  residual-in-residual dense block that improves the SISR performance.  
Furthermore, DeblurGAN~\cite{kupyn2018deblurgan} uses conditional GANs~\cite{mirza2014conditional} that allow constraining the generator to learn a pixel-to-pixel mapping~\cite{isola2017image} within the LR-HR domain. 
Recently, inspired by the success of CycleGAN~\cite{zhu2017unpaired} and DualGAN~\cite{yi2017dualgan}, Yuan \textit{et al.}~\cite{yuan2018unsupervised} proposed a cycle-in-cycle GAN-based model that can be trained using unpaired data. However, such unpaired training of GAN-based SISR models are prone to instability and often produce inconsistent results.

\subsection{SISR for Underwater Imagery}
SISR techniques for underwater imagery, on the other hand, are significantly less studied. 
As mentioned in the previous section, this is mostly due to the lack of large-scale datasets that capture the distribution of the unique distortions prevalent in underwater imagery. The existing datasets are only suitable for underwater object detection~\cite{islam2018understanding} and image enhancement~\cite{islam2019fast} tasks, as their image resolution is typically limited to $256\times256$, and they often contain synthetic images~\cite{fabbri2018enhancing}. Consequently, the performance and applicability of existing and novel SISR models for underwater imagery have not been explored in depth. 

Nevertheless, a few research attempts have been made for underwater SISR 
which primarily focus on reconstructing better quality underwater images from their noisy or blurred counterparts~\cite{chen2012model,fan2010application,yu2007system}. 
Other similar approaches have used SISR models to enhance underwater image sequence~\cite{quevedo2017underwater}, and to improve fish recognition performance~\cite{sun2016fish}.     
Although these models perform reasonably well for the respective applications, there is still significant room for improvement to match the SOTA performance. 
We attempt to address these aspects in this paper.


\begin{figure*}[ht]
	\centering
	\begin{subfigure}{0.98\textwidth}
		\centering
		\includegraphics[width=\linewidth]{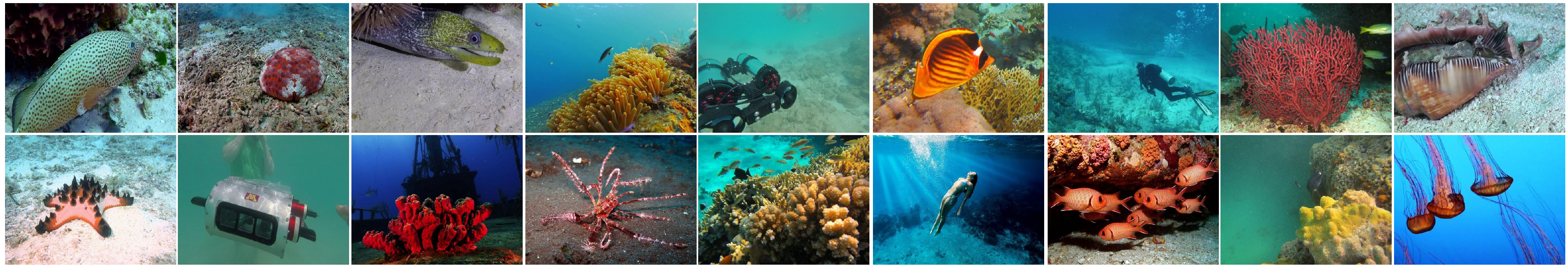} 
		\caption{A few instances sampled from the HR set; the HR images are of size $640\times480$.}
		\label{data_a}
	\end{subfigure}
	\vspace{2mm}
	
	\begin{subfigure}{0.75\textwidth} 
		\centering
		\includegraphics[width=0.98\linewidth]{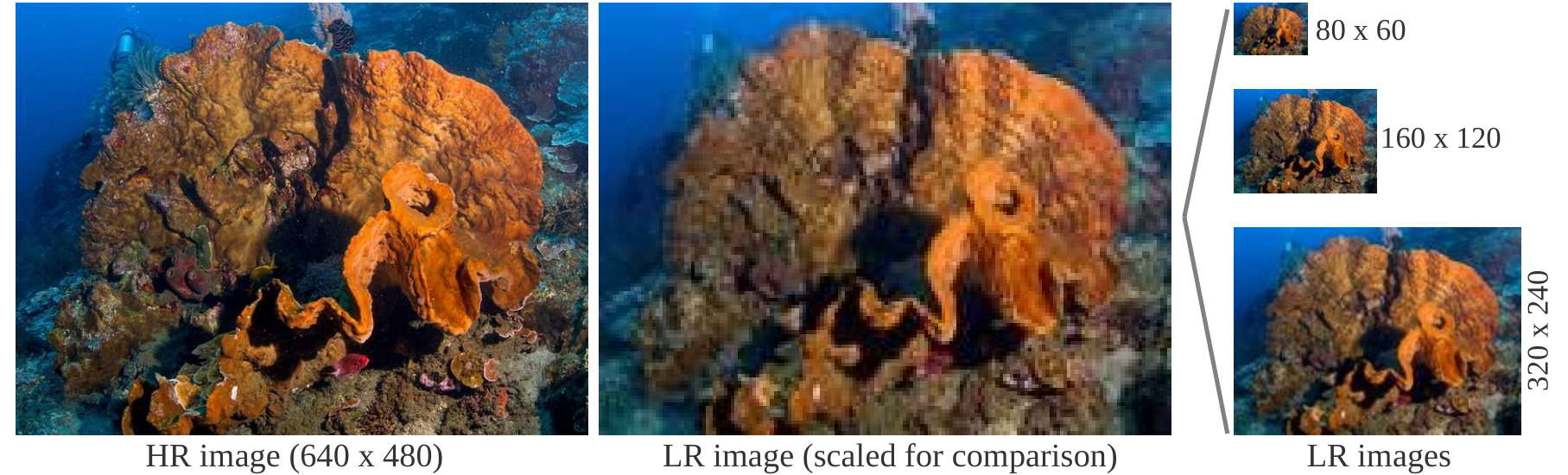} 
		\caption{A particular HR ground truth image and its corresponding LR images are shown.}
		\label{data_b}
	\end{subfigure}~
	\begin{subfigure}{0.24\textwidth} 
		\centering
		\includegraphics[width=0.98\linewidth]{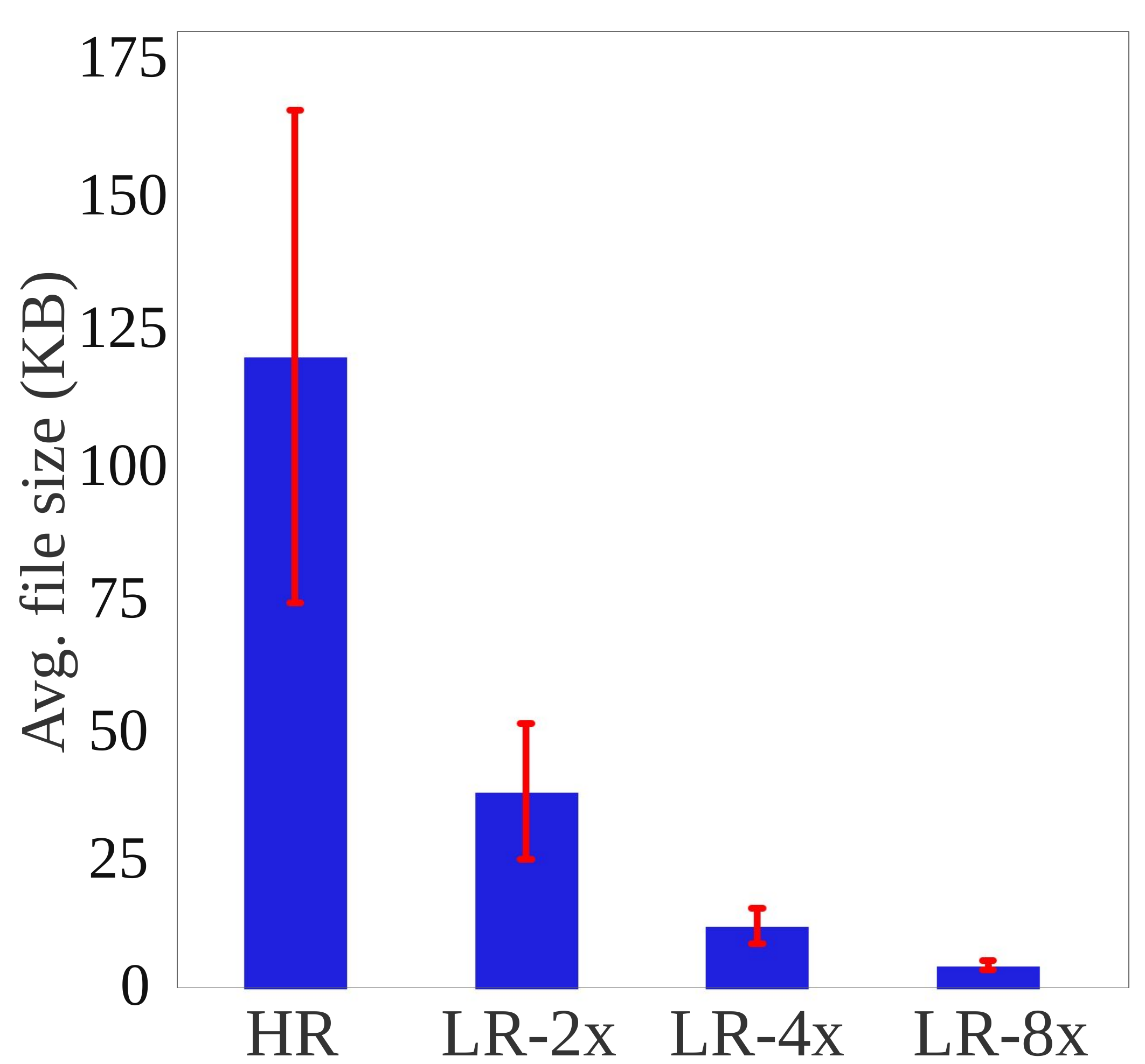} 
		\caption{Comparison of files sizes.}
		\label{data_c}
	\end{subfigure}
	\caption{The proposed USR-248 dataset has one HR set and three corresponding LR sets of images; hence, there are three possible combinations (\ie,  $2\times$, $4\times$ and $8\times$) for supervised training of SISR models.}
	\vspace{-2mm}
	\label{fig:data}
\end{figure*}

\section{USR-248 Dataset}
The USR-248 dataset contains a large collection of HR underwater images and their respective LR pairs. As mentioned earlier, there are three sets of LR images of size $80\times60$, $160\times120$, and $320\times240$; whereas, the HR images are of size $640\times480$. Each set has $1060$ RGB images for training and validation; another $248$ test images are provided for benchmark evaluation. A few sample images from the dataset are provided in Fig.~\ref{fig:data}.

\begin{figure}[hb]
	\centering
	\includegraphics[width=0.98\linewidth]{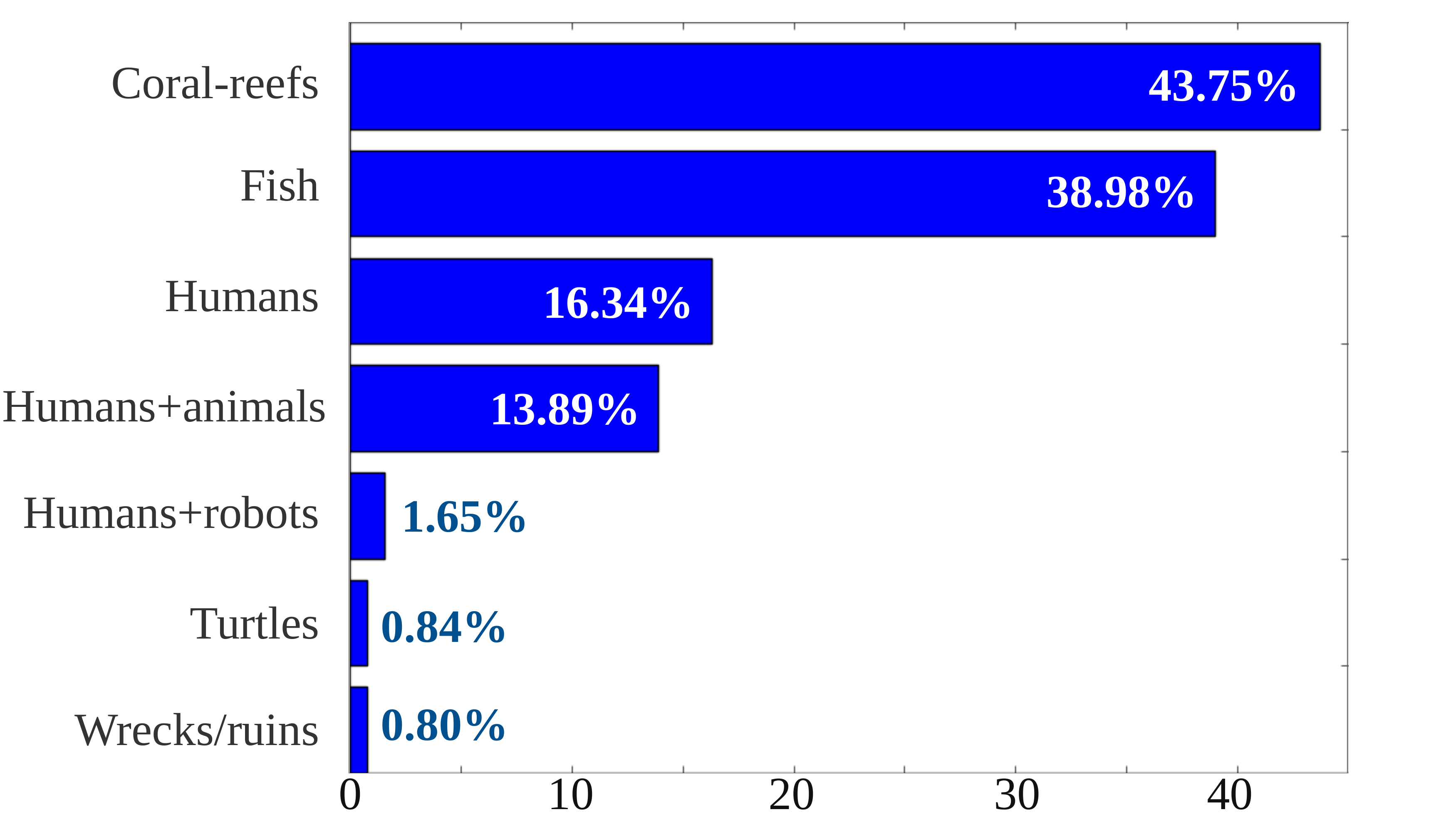} 
	\vspace{-1mm}
	\caption{Modality in the USR-248 dataset based on major objects of interest in the scene.}
	\label{fig:data_stat}
\end{figure}

To prepare the dataset, we collected HR underwater images: \textit{i)} during various oceanic explorations and field experiments, and \textit{ii)} from publicly available Flickr\textsuperscript{TM} images and YouTube\textsuperscript{TM} videos. The field experiments are performed in a number of different locations over a diverse set of visibility conditions. 
Multiple GoPros~\cite{gopro}, Aqua AUV's uEye cameras~\cite{dudek2007aqua}, low-light USB cameras \cite{lowlight}, and Trident ROV's HD camera~\cite{trident} are used to collect HR images during the experiments. 
We also compiled HR underwater images containing natural scenes from Flickr\textsuperscript{TM}, YouTube\textsuperscript{TM}, and other online resources\footnote{Detailed information and credits for the online media resources can be found in Appendix I.}. We avoided multiple instances of similar scenes and made sure they contain different objects of interest (\eg, coral reefs, fish, divers, wrecks/ruins, etc.) in a variety of backgrounds. Fig.~\ref{fig:data_stat} shows the modality in the data in terms of object categories. Once the HR images are selected and resized to $640\times480$, three sets of LR images are generated by compressing and then gradually downsizing the images to $320\times240$, $160\times120$, and $80\times60$; a comparison of the average file sizes for these image sets are shown in Fig.~\ref{data_c}. Overall, USR-248 provides large-scale paired data for training $2\times$, $4\times$, and $8\times$ underwater SISR models. It also includes the respective validation and test sets that are used to evaluate our proposed model.    


\section{SRDRM and SRDRM-GAN Model}

\subsection{Deep Residual Multiplier (DRM)}
\begin{figure*}[t]
	\centering
	\begin{subfigure}{0.43\textwidth}
		\includegraphics[width=0.95\linewidth]{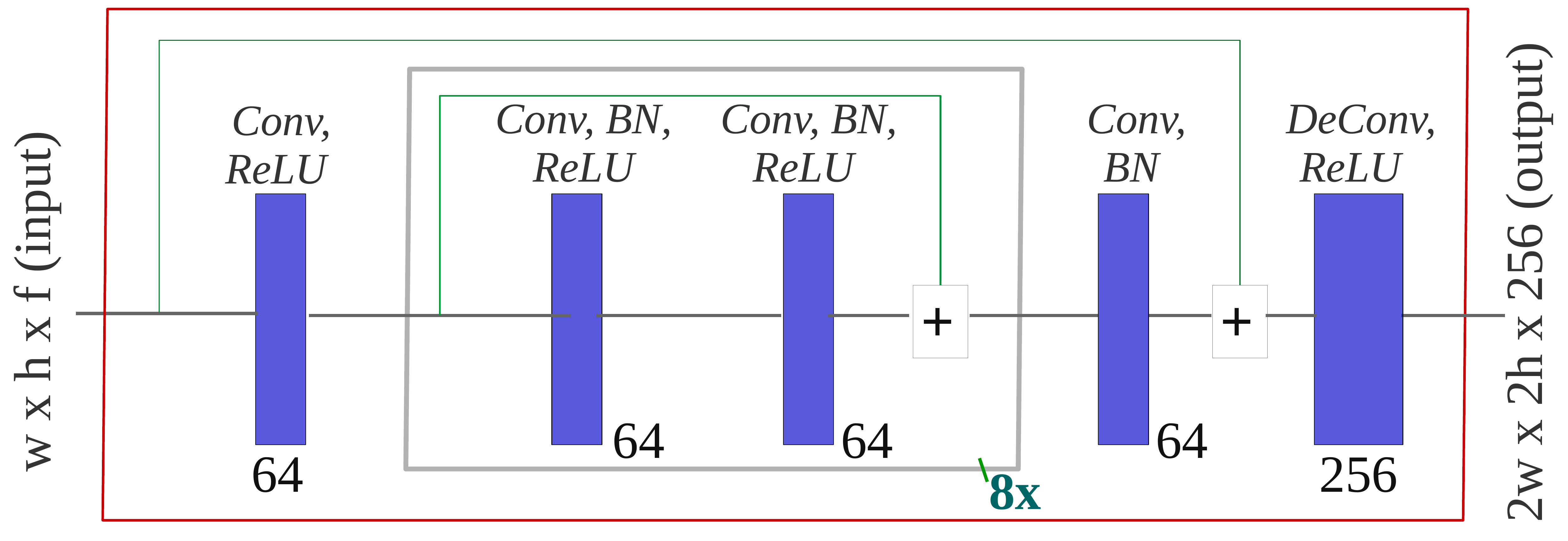}
		\caption{Architecture of a deep residual multiplier (DRM) block}
		\label{fig:model_drm}
	\end{subfigure}~
	\begin{subfigure}{0.48\textwidth}
		\includegraphics[width=0.95\linewidth]{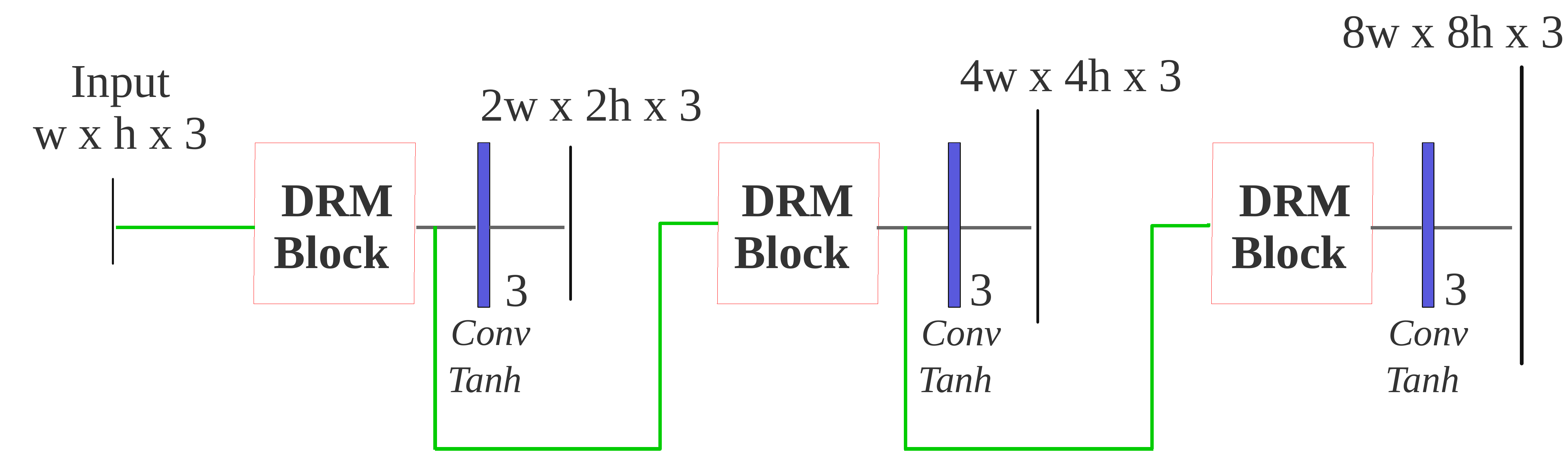}
		\caption{Generator: SRDRM model with multiple DRM blocks}
		\label{fig:model_srdrm}
	\end{subfigure}
	
	\vspace{1mm}
	\begin{subfigure}{0.92\textwidth}
		\includegraphics[width=0.95\linewidth]{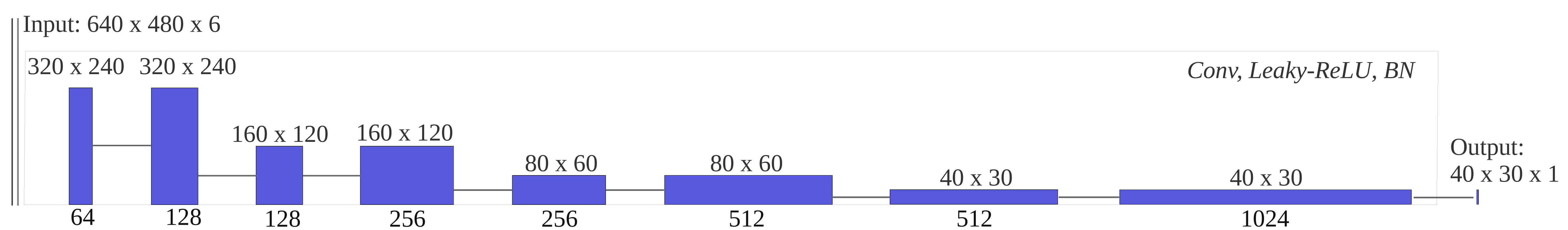}
		\caption{Discriminator: a Markovian PatchGAN~\cite{isola2017image} with nine layers and a patch-size of $40\times30$}
		\label{fig:model_srdrmgan}
	\end{subfigure}
	\caption{Network architecture of the proposed model.}
	\vspace{-2mm}
	\label{fig:model}
\end{figure*}

The core element of the proposed model is a fully-convolutional deep residual block, designed to learn $2\times$ interpolation in the RGB image space. We denote this building block as Deep Residual Multiplier (DRM) as it scales the input features' spatial dimensions by a factor of two. As illustrated in Figure~\ref{fig:model_drm}, DRM consists of a convolutional ({\tt conv}) layer, followed by $8$ repeated residual layers, then another {\tt conv} layer, and finally a de-convolutional (\ie, {\tt deconv}) layer for up-scaling.
Each of the repeated residual layers (consisting of two {\tt conv} layers) is designed by following the principles outlined in the EDSR model~\cite{lim2017enhanced}. 
Several choices of hyper-parameters, \eg, the number of filters in each layer, the use of {\tt ReLU} non-linearity~\cite{nair2010rectified}, and/or Batch Normalization (BN)~\cite{ioffe2015batch} are annotated in Fig.~\ref{fig:model_drm}. As a whole, DRM is a $10$ layer residual network that learns to scale up the spatial dimension of input features by a factor of two. It uses a series of 2D convolutions of size $3\times3$ (in repeated residual block) and $4\times4$ (in the rest of the network) to learn this spatial interpolation from paired training data.

\subsection{SRDRM Architecture}
As Fig.~\ref{fig:model_srdrm} demonstrates, the SRDRM makes use of $n \in \{1, 2, 3\}$ DRM blocks in order to learn to generate $2^n\times$ HR outputs. An additional {\tt conv} layer with {\tt tanh} non-linearity~\cite{raiko2012deep} is added after the final DRM block in order to reshape the output features to the desired shape. Specifically, it generates a $2^nw \times 2^nh \times3$ output for an input of size $w\times h \times3$. 


\subsection{SRDRM-GAN Architecture}
For adversarial training, we use the same SRDRM model as the \textit{generator} and employ a Markovian PatchGAN~\cite{isola2017image}-based model for the \textit{discriminator}. 
As illustrated by Fig.~\ref{fig:model_srdrmgan}, nine {\tt conv} layers are used to transform a $640\times480\times6$ input (real and generated image) to a $40\times30\times1$ output that represents the averaged \textit{validity} responses of the discriminator. At each layer, $3\times3$ convolutional filters are used with a stride size of $2$, followed by a {\tt Leaky-ReLU} non-linearity~\cite{maas2013rectifier} and BN. 
Although traditionally PatchGANs use $70\times70$ patches~\cite{isola2017image,yi2017dualgan}, we use a patch-size of $40\times30$ as our input/output image-shapes are of $4$:$3$.

\subsection{Objective Function Formulation}
At first, we define the SISR problem as learning a function or mapping $G:\{X\} \rightarrow Y$, where $X$ ($Y$) represents the LR (HR) image domain.  
Then, we formulate an objective function that evaluates the following properties of $G(X)$ compared to $Y$:   

 \textit{1)} \textbf{Global similarity and perceptual loss}: existing methods have shown that adding an $L_1$ ($L_2$) loss to the objective function enables the generator to learn to sample from a globally similar space in an $L_1$ ($L_2$) sense~\cite{isola2017image}. In our implementation, we measure the \textit{global similarity} loss as:      	$\mathcal{L}_{2}(G) = \mathbb{E}_{X,Y} \big[\big|\big|Y-G(X)\big|\big|_2\big]$. Additionally, as suggested in~\cite{compuphase}, we define a \textit{perceptual loss} function based on the per-channel disparity between $G(X)$ and $Y$ as:  
        \[
        	\mathcal{L}_{P}(G) = \mathbb{E}_{X,Y} \big[\big|\big| (512+\bar{\mathbf r})\mathbf r^2+4\mathbf g^2+(767-\bar{\mathbf r})\mathbf b^2  \big|\big|_2\big].
        \]
    Here, $\mathbf r$, $\mathbf g$, and $\mathbf b$ denote the normalized numeric differences of the red, green, and blue channels between $G(X)$ and $Y$, respectively; whereas $\bar{\mathbf r}$ is the mean of their red channels.  

    \textit{2)} \textbf{Image content loss}: being inspired by the success of existing SISR models~\cite{yang2019deep}, we also formulate the \textit{content loss} as:     
            \[
        	\mathcal{L}_{C}(G) = \mathbb{E}_{X,Y} \big[\big|\big|\Phi (Y)-\Phi (G(X)) \big|\big|_2\big].
            \]
    Here, the function $\Phi(\cdot)$ denotes the high-level features extracted by the {\tt block5\_conv4} layer of a pre-trained {\tt VGG-19} network.

Finally, we formulate the multi-modal objective function for the generator as: 
$
	\mathcal{L_G}(G) = \lambda_c \mathcal{L}_{C}(G) + \lambda_p \mathcal{L}_{P}(G)+\lambda_2 \mathcal{L}_{2}(G). 
$
Here, $\lambda_c$, $\lambda_p$, and $\lambda_2$ are scalars that are empirically tuned as hyper-parameters. Therefore, the generator $G$ needs to solve the following minimization problem:
\begin{equation} \label{eq:g_final}
\centering
	G^* = \argmin\limits_{G} \mathcal{L_G}(G). 
\end{equation}

On the other hand, adversarial training requires a two-player min-max game~\cite{goodfellow2014generative} between the generator $G$ and discriminator $D$, which is expressed as:   
\begin{equation} 
\centering
  \footnotesize
	\mathcal{L}(G,D) = \mathbb{E}_{X,Y} \big[\log D(Y)\big]+ \mathbb{E}_{X,Y} \big[\log (1-D(X, G(X)))\big].
\end{equation}
Here, the generator tries to minimize $\mathcal{L}(G,D)$ while the discriminator tries to maximize it. Therefore, the optimization problem for adversarial training becomes: 
\begin{equation} \label{eq:gan_final}
\centering
	G^* = \argmin\limits_{G}\max\limits_{D} \mathcal{L}_{GAN}(G,D)+ \mathcal{L_G}(G).
\end{equation}


\subsection{Implementation}
We use TensorFlow libraries~\cite{abadi2016tensorflow} to implement the proposed SRDRM and SRDRM-GAN models.
We trained both the models on the USR-248 dataset up to $20$ epochs with a batch-size of $4$, using two NVIDIA\textsuperscript{TM} GeForce GTX 1080 graphics cards.
We also implement a number of SOTA generative and adversarial models for performance comparison in the same setup. Specifically, we consider three generative models named SRCNN~\cite{dong2015image}, SRResNet~\cite{ledig2017photo,yang2019deep}, and DSRCNN~\cite{mao2016image}, and three adversarial models named SRGAN~\cite{ledig2017photo}, ESRGAN~\cite{wang2018esrgan}, and EDSRGAN~\cite{lim2017enhanced}. 
We already provided a brief discussion on the SOTA SISR models in  Section~\ref{related_work}.
Next, we present the experimental results based on qualitative analysis and quantitative evaluations in terms of standard metrics.

\section{Experimental Results}

\subsection{Qualitative Evaluations}
\begin{figure}[b]
\vspace{-2mm}
    \centering
    \includegraphics[width=0.98\linewidth]{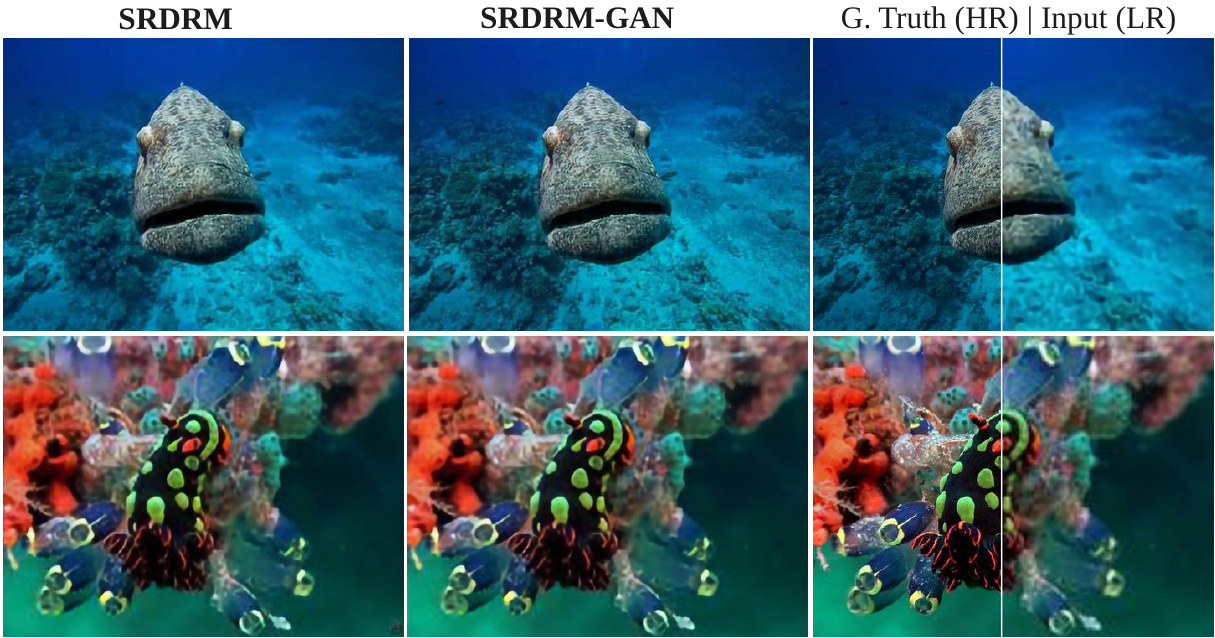}
    \vspace{-1mm}
    \caption{Color consistency and sharpness of the generated $4\times$ HR images compared to the respective ground truth.}
    \label{fig:col}
\end{figure}

\begin{figure}[t]
    \centering
    \includegraphics[width=0.98\linewidth]{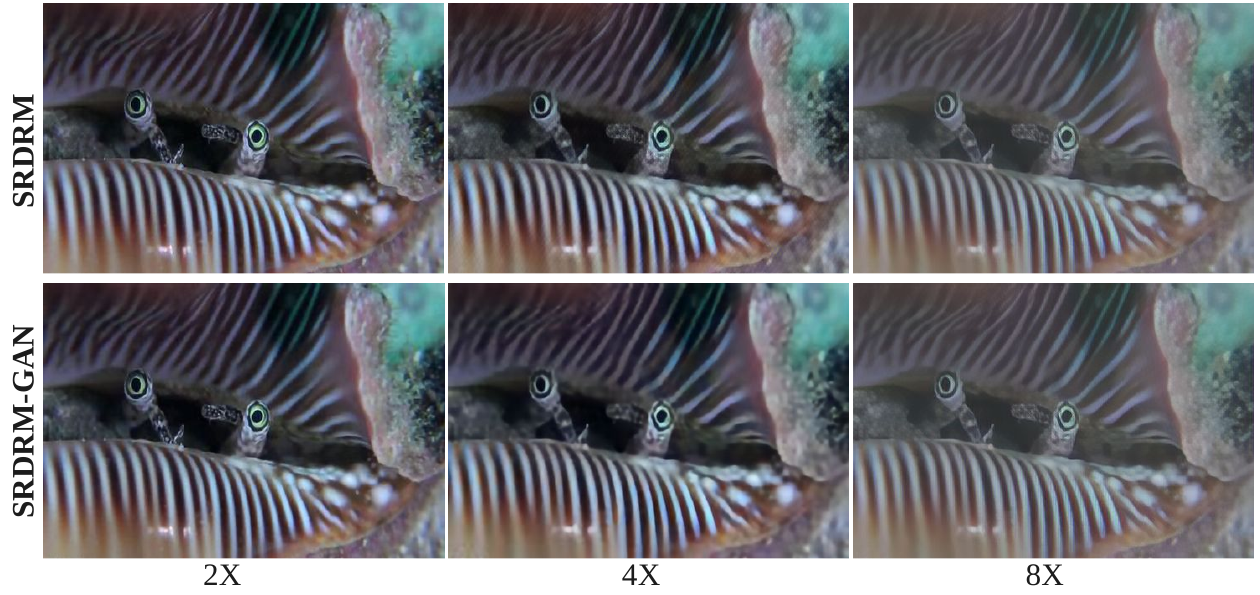}
    \vspace{-1mm}
    \caption{Global contrast and texture recovery by SRDRM and SRDRM-GAN for $2\times$, $4\times$, and $8\times$ SISR.}
    \label{fig:con}
\end{figure}

At first, we analyze the sharpness and color consistency in the generated images of SRDRM and SRDRM-GAN. As Fig.~\ref{fig:col} suggests, both models generate images that are comparable to the ground truth for $4\times$ SISR. We observe even better results for $2\times$ SISR, as it is a relatively less challenging problem. We demonstrate this relative performance margins at various scales in Fig.~\ref{fig:con}. This comparison shows that the global contrast and texture is mostly recovered in the $2\times$ and $4\times$ HR images generated by SRDRM and SRDRM-GAN. On the other hand, the $8\times$ HR images miss the finer details and lack the sharpness in high-texture regions. The state-of-the-art SISR models have also reported such difficulties beyond the $4\times$ scale~\cite{yang2019deep}. 

\begin{figure*}[t]
    \centering
    \includegraphics[width=0.98\linewidth]{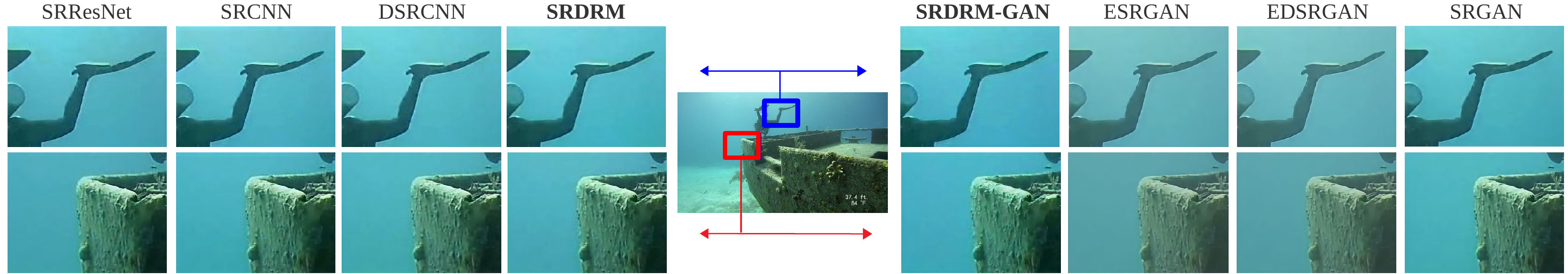}  \\
    \vspace{1mm}
    \includegraphics[width=0.98\linewidth]{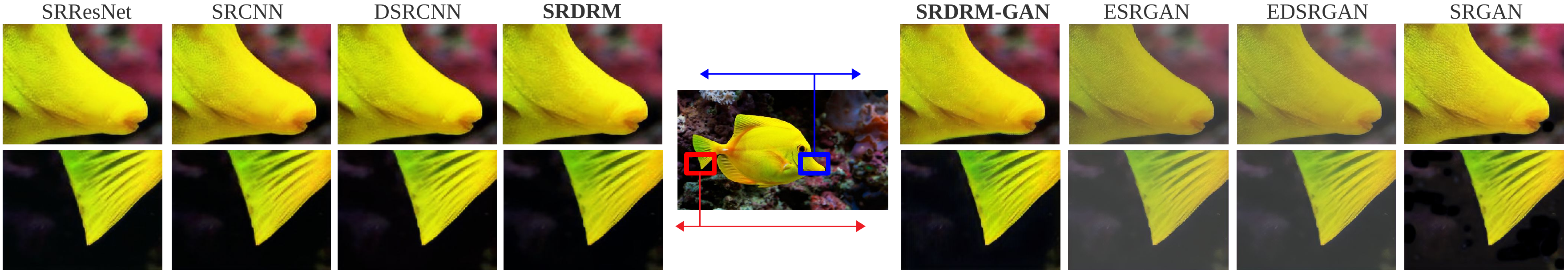}\\
    \vspace{1mm}
    \includegraphics[width=0.98\linewidth]{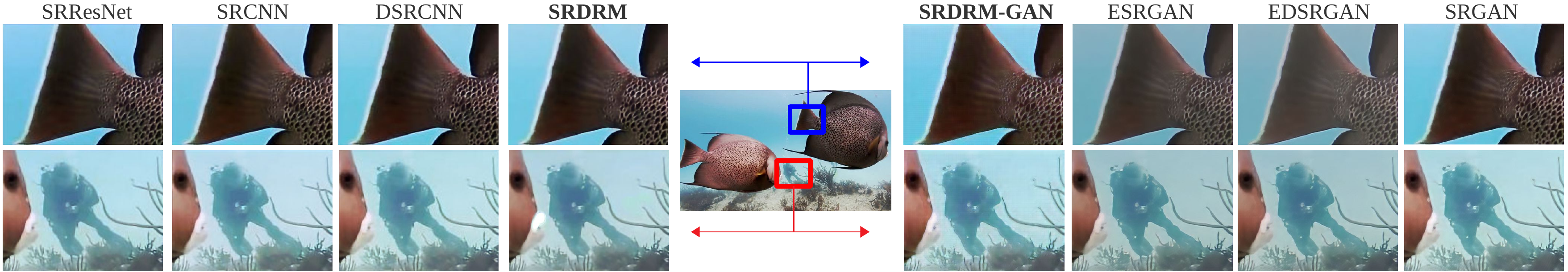}\\
    \vspace{-1mm}
    \caption{Qualitative performance comparison of SRDRM and SRDRM-GAN with SRCNN~\cite{dong2015image}, SRResNet~\cite{ledig2017photo,yang2019deep},  DSRCNN~\cite{mao2016image}, 
    SRGAN~\cite{ledig2017photo}, 
    ESRGAN~\cite{wang2018esrgan}, and EDSRGAN~\cite{lim2017enhanced}. (Best viewed at $400\%$ zoom)   
    }
    \vspace{-3mm}
    \label{fig:comp}
\end{figure*}

Next, in Fig.~\ref{fig:comp}, we provide a qualitative performance comparison with the state-of-the-art models for $4\times$ SISR. 
We select multiple $160\times120$ patches on the test images containing interesting textures and objects in contrasting background. Then, we apply all the SISR models (trained on $4\times$ USR-248 data) to generate respective HR images of size $640\times480$.      
In the evaluation, we observe that SRDRM performs at least as well as and often better compared to the generative models, \ie, SRResNet, SRCNN, and DSRCNN. Moreover, SRResNet and SRGAN are prone to inconsistent coloring and over-saturation in bright regions. 
On the other hand, ESRGAN and EDSRGAN often fail to restore the sharpness and global contrast. 
Furthermore, SRDRM-GAN generates sharper images and does a better texture recovery than SRDRM (and other generative models) in general. We postulate that the PatchGAN-based discriminator contributes to this, as it forces the generator to learn high-frequency local texture and style information~\cite{isola2017image}.


\subsection{Quantitative Evaluation}
We consider two standard metrics~\cite{hore2010image,islam2019fast} named Peak Signal-to-Noise Ratio (PSNR) and Structural Similarity (SSIM) in order to quantitatively compare the SISR models' performances. 
The PSNR approximates the reconstruction quality of generated images compared to their respective ground truth, whereas the SSIM~\cite{wang2004image} compares the image patches based on three properties: luminance, contrast, and structure.
In addition, we consider Underwater Image Quality Measure (UIQM)~\cite{panetta2016human}, which 
quantifies underwater image colorfulness, sharpness, and contrast.
We evaluate all the SISR models on USR-248 test images, and compare their performance in Table~\ref{tab:psnr_ssim}. 
The results indicate that SRDRM-GAN, SRDRM, SRGAN, and SRResNet produce comparable values for PSNR and SSIM, and perform better than other models. SRDRM and SRDRM-GAN also produce higher UIQM scores than other models in comparison. These statistics are consistent with our qualitative analysis.

\begin{table}[t]
\centering
\caption{Comparison of average PSNR, SSIM, and UIQM scores for $2\times$/$4\times$/$8\times$ SISR on USR-248 test set.}
\scriptsize
\vspace{-1mm}
\begin{tabular}{l|c|c|c}
  \hline
   & $PSNR$ & $SSIM$ & $UIQM$ \\
  \textbf{Model}    & $\big(G(\mathbf{x}),\mathbf{y}\big)$ & $\big(G(\mathbf{x}),\mathbf{y}\big)$ & $\big(G(\mathbf{x})\big)$ \\
  \hline \hline
  SRResNet & $25.98$/$24.15$/$19.26$ & $0.72$/$0.66$/$0.55$ & $2.68$/$2.23$/$1.95$ \\ \hline
  SRCNN & $26.81$/$23.38$/$19.97$ & $0.76$/$0.67$/$0.57$ & $2.74$/$2.38$/$2.01$ \\ \hline
  DSRCNN & $27.14$/$23.61$/$20.14$ & $0.77$/$0.67$/$0.56$ & $2.71$/$2.36$/$2.04$ \\ \hline
  \textbf{SRDRM} & $28.36$/$24.64$/$21.20$ & $0.80$/$0.68$/$0.60$ & $2.78$/$2.46$/$2.18$ \\ \hline
  \textbf{SRDRM-GAN} & $28.55$/$24.62$/$20.25$ & $0.81$/$0.69$/$0.61$ & $2.77$/$2.48$/$2.17$ \\ \hline
  ESRGAN & $26.66$/$23.79$/$19.75$ & $0.75$/$0.66$/$0.58$ & $2.70$/$2.38$/$2.05$ \\ \hline
  EDSRGAN & $27.12$/$21.65$/$19.87$ & $0.77$/$0.65$/$0.58$ & $2.67$/$2.40$/$2.12$ \\ \hline
  SRGAN & $28.05$/$24.76$/$20.14$ & $0.78$/$0.69$/$0.60$ & $2.74$/$2.42$/$2.10$ \\ \hline
\end{tabular}
\vspace{-1mm}
\label{tab:psnr_ssim}
\end{table}

\subsection{Practical Feasibility}
The qualitative and quantitative results suggest that SRDRM and SRDRM-GAN provide good quality HR visualizations for LR image patches, which is potentially useful in tracking fast-moving targets, attention modeling, and detailed understanding of underwater scenes. Therefore, AUVs and ROVs can use this to \emph{zoom in} a particular region of interest (RoI) for detailed and improved visual perception. 
One operational consideration for using such deep learning-based models in embedded robotic platforms is the computational complexity. As we demonstrate in Table~\ref{tab:time}, the memory requirement for the proposed model is only $3.5$-$12$ MB and it runs at $4$-$7$ fps on NVIDIA\textsuperscript{TM} Jetson TX2. 
Therefore, it essentially takes about $140$-$246$ milliseconds for a robot to take a closer look at a LR RoI.  
These results validate the feasibility of using the proposed model for improving real-time perception of visually-guided underwater robots.

\begin{table}[t]
\centering
\caption{Run-time and memory requirement of SRDRM (same as SRDRM-GAN) on NVIDIA\textsuperscript{TM} Jetson TX2 (optimized graph).}
\footnotesize
\begin{tabular}{l||c|c|c}
  \hline
  \textbf{Model} & $2\times$  & $4\times$  & $8\times$  \\ \hline \hline
  Inference-time (ms) &  $140.6$ ms  &  $145.7$ ms & $245.7$ ms \\
  Frames per second (fps)  &  $7.11$ fps &  $6.86$  fps & $4.07$  fps \\
  \hline
  Model-size &  $3.5$ MB &  $8$ MB & $12$ MB \\ \hline
\end{tabular}
\vspace{-3mm}
\label{tab:time}
\end{table}%
\section{Conclusion}
In this paper, we present a fully-convolutional deep residual network-based model for underwater image super-resolution at $2\times$, $4\times$, and $8\times$ scales. 
We also provide generative and adversarial training pipelines driven by a multi-modal objective function, which is designed to evaluate image quality based on its content, color, and texture information. 
In addition, we present a large-scale dataset named USR-248 which contains paired underwater images of various resolutions 
for supervised training of SISR models. 
Furthermore, we perform thorough qualitative and quantitative evaluations which suggest that the proposed model can learn to restore image qualities 
at a higher resolution for an improved visual perception. 
In the future, we seek to improve its performance for $8\times$ SISR, and plan to further investigate its applicability in other underwater robotic applications.


{\small
\bibliographystyle{ieee}
\bibliography{Refs}
}

\section*{Appendix I: Credits for Media Resources}
{\small
\begin{enumerate}   
    
    \item Wallpapercave.com. Sea-turtle. 2018. (Wallpapercave):
    \url{https://wallpapercave.com/w/wp4430950}.
    


    
    
    
    
    \item Simon Gingins. Potato cod (Epinephelus tukula) - Great Barrier Reef - Australia. 2014. (Flickr):
    \url{https://www.flickr.com/photos/simongingins/15574452614/}
    
    \item Cat Trumpet. 2 Hours of Beautiful Coral Reef Fish, Relaxing Ocean Fish, 1080p HD. 2016. (YouTube): \\ \url{https://youtu.be/cC9r0jHF-Fw}.

    \item Nature Relaxation Films. 3 Hours of Stunning Underwater Footage, French Polynesia, Indonesia. 2018. (YouTube): \\
    \url{https://youtu.be/eSRj847AY8U}.

    \item Calm Cove Club - Relaxing Videos. 4K Beautiful Ocean Clown Fish Turtle Aquarium. 2017 (YouTube): \\ \url{https://youtu.be/DP4QDNm6f4Q}.
    
    \item Scubasnap.com. 4K Underwater at Stuart Cove's, 2014 (YouTube):  
    \url{https://youtu.be/kiWfG31YbXo}.

    \item 4.000 PIXELS. Beautiful Underwater Nature. 2017 (YouTube): \url{https://youtu.be/1-Cn0b1MKrM}.
    
    \item Magnus Ryan Diving. SCUBA Diving Egypt Red Sea. 2017 (YouTube):
    \url{https://youtu.be/CaLfMHl3M2o}.
    
    \item Soothing Relaxation. Sleep Music in Underwater Paradise. 2017 (YouTube): \url{https://youtu.be/OVct34NUk3U}.

    
    
    \item TheSilentWatcher. 4K Coral World-Tropical Reef. 2018. (YouTube):
    \url{https://youtu.be/uyb0wW0ln_g}.
    
    \item Awesome Video. 4K- The Most Beautiful Coral Reefs and Undersea Creature on Earth. 2017. (YouTube): \\
    \url{https://youtu.be/nvq_lvC1MRY}.
    
    \item Earth Touch. Celebrating World Oceans Day in 4K. 2015. (YouTube):
    \url{https://youtu.be/IXxfIMNgMJA}.
    
    \item BBC Earth. Deep Ocean: Relaxing Oceanscapes. 2018. (YouTube):
    \url{https://youtu.be/t_S_cN2re4g}.
    
    \item Alegra Chetti. Let's Go Under the Sea I Underwater Shark Footage I Relaxing Underwater Scene. 2016. (YouTube):
    \url{https://youtu.be/rQB-f5BHn5M}.
    
    \item Underwater 3D Channel- Barry Chall Films. Planet Earth, The Undersea World (4K). 2018. (YouTube): \\
    \url{https://youtu.be/567vaK3BKbo}.
    
    \item Undersea Productions. ``ReefScapes: Nature's Aquarium" Ambient Underwater Relaxing Natural Coral Reefs and Ocean Nature. 2009. (YouTube): \\
    \url{https://youtu.be/muYaOHfP038}.
    
    \item BBC Earth. The Coral Reef: 10 Hours of Relaxing Oceanscapes. 2018. (YouTube): \\
    \url{https://youtu.be/nMAzchVWTis}.
    
    \item Robby Michaelle. Scuba Diving the Great Barrier Reef Red Sea Egypt Tiran. 2014. (YouTube): \\
    \url{https://youtu.be/b7BEAsyPgHM}.
    
    \item Bubble Vision. Diving in Bali. 2012. (YouTube): \\
    \url{https://youtu.be/uCRBxtQ55_Y}. 
    
    \item Vic Stefanu - Amazing World Videos. EXPLORING The GREAT BARRIER REEF, fantastic UNDERWATER VIDEOS (Australia). 2015. (YouTube): \\
    \url{https://youtu.be/stMzgmPlQQM}.
    
    \item Our Coral Reef. Breathtaking Dive in Raja Ampat, West Papua, Indonesia Coral Reef. 2018. (YouTube): \\
    \url{https://youtu.be/i4ZSMDWNXTg}.
    
    \item GoPro. GoPro Awards: Great Barrier Reef with Fusion Overcapture in 4K. 2018. (YouTube): \\
    \url{https://youtu.be/OAmBkfn62dY}.
    
    \item GoPro. GoPro: Freediving with Tiger Sharks in 4K. 2017. (YouTube): 
    \url{https://youtu.be/Zy3kdMFvxUU}.
    \item TFIL. SCUBA DIVING WITH SHARKS!. 2017. (YouTube):
    \url{https://youtu.be/v8eSPf4RzTU}.
    
    \item Vins and Annette Singh. Stunning salt Water Fishes in a Marine Aquarium. 2019. (YouTube): \\
    \url{https://youtu.be/CWzXL6a4KGM}.
    
    \item Akouris. H.M.Submarine Perseus. 2014. (YouTube):
    \url{https://youtu.be/4-oP0sX723k}.
    
    \item Gung Ho Vids. U.S. Navy Divers View An Underwater Wreck. 2014. (YouTube): \\
    \url{https://youtu.be/1qfRQRUMnXY}.
    
    \item Martcerv. Truk lagoon deep wrecks, GoPro black with SRP tray and lights. 2013. (YouTube): \\
    \url{https://youtu.be/0uD-nCN03s8}.
    
    \item Dmireiy. Shipwreck Diving, Nassau Bahamas. 2012. (YouTube):
    \url{https://youtu.be/CIQI3isddbE}.
    
    \item Frank Lame. diving WWII Wrecks around Palau. 2010. (YouTube):
    \url{https://youtu.be/vcI63XQsNlI}.
    
    \item Stevanurk. Wreck Dives Malta. 2014. (YouTube): \\
    \url{https://youtu.be/IZFuOIwEBH8}.
    
    \item Stevanurk. Diving Malta, Gozo and Comino 2015 Wrecks Caves. 2015. (YouTube): \\
    \url{https://youtu.be/NrDDjnij7sA}.
    
    \item Octavio velazquez lozano. SHIPWRECK Scuba Diving BAHAMAS. 2017. (YouTube):\\
    \url{https://youtu.be/4ovFPCEw4Qk}.
    
    \item Drew Kaplan. SCUBA Diving The Sunken Ancient Roman City Of Baiae, the Underwater Pompeii. 2018. (YouTube): \\
    \url{https://youtu.be/8RmJ3jzrwH8}.
    
    \item Octavio velazquez lozano. LIBERTY SHIPWRECK scuba dive destin florida. 2017. (YouTube): \\
    \url{https://youtu.be/DHuHZdVWONk}.
    
    \item Blue Robotics. BlueROV2 Dive: Hawaiian Open Water. 2016. (YouTube): \\
    \url{https://youtu.be/574jPVEk7mo}.
    
    \item JerryRigEverything. Exploring a Plane Wreck - UNDER WATER!. 2018. (YouTube): \\
    \url{https://youtu.be/0-sZVJbUzqo}.
    
    \item Rovrobotsubmariner. Home-built Underwater Robot ROV in Action!. 2010. (YouTube): \\
    \url{https://youtu.be/khLEyyf3Ci8}.
    
    \item Oded Ezra. Eca-Robotics H800 ROV. 2016. (YouTube): \\
    \url{https://youtu.be/Yafq9c7cqgE}.
    
    \item Geneinno Tech. Titan Diving Drone. 2019. (YouTube): \\
    \url{https://youtu.be/h7Bn4MxkFxs}.
    
    \item Scubo. Scubo - Agile Multifunctional Underwater Robot - ETH Zurich. 2016. (YouTube): \\
    \url{https://youtu.be/-g2O8e1j3fw}.
    
    \item Learning with Shedd. Student-built Underwater Robot at Shedd ROV Club Event. 2017. (YouTube): \\
    \url{https://youtu.be/y3dn8snT8os}.
    
    \item HMU-CSRL. SQUIDBOT sea trials. 2015. (YouTube): \\
    \url{https://youtu.be/0iDBF23gI6I}. 
    
    \item MobileRobots. Aqua2 Underwater Robot Navigates in a Coral Reef - Barbados. 2012. (YouTube): \\
    \url{https://youtu.be/jC-AmPfInwU}.
    
    \item Daniela Rus. underwater robot. 2015. (YouTube): \\
    \url{https://youtu.be/neLu0ZGuXPM}.
    
    \item JohnFardoulis. Sirius - Underwater Robot, Mapping. 2014. (YouTube): 
    \url{https://youtu.be/fXxVcucOPrs}.

\end{enumerate}

}

\end{document}